\documentclass[twocolumn,showpacs,preprintnumbers]{revtex4}
\usepackage{graphicx}% Include figure files
\usepackage{dcolumn}% Align table columns on decimal point
\usepackage{bm}% bold math

\def\lesssim{\ \raise.3ex\hbox{$<$}\kern-0.8em\lower.7ex\hbox{$\sim$}\ }
\def\gesim{\ \raise.3ex\hbox{$>$}\kern-0.8em\lower.7ex\hbox{$\sim$}\ }
\font\scripti=cmmi7
\font\scriptscripti=cmmi5
\def\sib#1{\setbox0 = \hbox{\scripti #1}
  \kern-.02em\copy0\kern-\wd0
  \kern.04em\box0} % script italic bold 
\def\ssib#1{\setbox0 = \hbox{\scriptscripti #1}
  \kern-.02em\copy0\kern-\wd0
  \kern.04em\box0} % scriptscript italic bold

\font\tenib=cmmib10 % italic bold for math
\skewchar\tenib='177 \skewchar\tenib='177 \skewchar\tenib='177
\def\pbold#1{\setbox0 = \hbox{$ #1 $}
  \kern-.022em\copy0\kern-\wd0
  \kern.011em\copy0\kern-\wd0
  \kern.011em\copy0\kern-\wd0
  \kern.011em\copy0\kern-\wd0
  \kern.011em\box0} % poorman's bold 

\begin{document}
%\preprint{APS/123-QED}

\title{Fermi excitations in a trapped atomic Fermi gas with a molecular Bose condensate}
\author{Y. Ohashi}
\affiliation{Institute of Physics, University of Tsukuba, Tsukuba, Ibaraki 305, Japan}
\author{A. Griffin}
\affiliation{Department of Physics, University of Toronto, Toronto, Ontario, Canada M5S 1A7}
\date{\today}
\begin{abstract}
We discuss the effect of a molecular Bose condensate on the energy
of Fermi excitations in a trapped
two-component atomic Fermi
gas. The single-particle Green's functions can be approximated by the
well-known BCS form, in both the BCS (Cooper pairs) and BEC (Feshbach
resonance molecules) domains. The composite Bose order parameter ${\tilde
  \Delta}$ describing bound states of two atoms and the Fermi chemical
potential $\mu$ are calculated self-consistently. In the
BEC regime characterized by $\mu<0$, the Fermi
quasiparticle energy gap is given by $\sqrt{\mu^2+{\tilde \Delta}^2}$,
instead of $|{\tilde \Delta}|$ in the BCS region, where $\mu>0$.
This shows up in the characteristic energy of atoms from
dissociated molecules.
\end{abstract}

\pacs{03.75.Ss, 05.30.Jp, 74.20.Mn}
\maketitle

%%%%%%%%%%%%%%%%%%%%%%%%%%%%%%%%%%%%%%%%%%%%%%%%%%%%%%%%%%%%%%%%%%%%%%%%%%%%%%%

There has been increasing interest in the BCS-BEC
crossover in a two-component trapped atomic Fermi
gas\cite{1Ohashi1,3Ohashi3,Milstein}. This recent theoretical work has 
extended the classic work in
superconductors\cite{4Leggett,5Nozieres,Melo,Randeria} to include a Feshbach
resonance\cite{7Holland,8Timmermans}. This allows one to both
increase the attractive interaction\cite{7Holland,8Timmermans} by
working above
the resonance as well as to include the effect of long-lived dimer molecules just
below the resonance. Several
groups\cite{9Jin,10Ketterle,11Jochim} have presented evidence for
a molecular Bose condensate in a two-component Fermi gas in
the region $a^{\rm 2b}_s>0$, where $a^{\rm 2b}_s$ is the two-body 
$s$-wave scattering length. 
Recent theories\cite{1Ohashi1,3Ohashi3,Milstein}
clearly show that a unified description of both the BEC and BCS limits
can be given, in terms of a {\it composite} order parameter which involves
a Cooper pair condensate and a molecular
Bose condensate associated with the Feshbach resonance. 
They are two limits of a superfluid Fermi gas, 
both arising from a Bose condensate of bound
states. The only difference lies in the origin of these bound states.
\par
This similarity has been noted in the study of the collective modes of
this condensate in the BCS and BEC limits\cite{Menotti,Baranov} as
well as in
the crossover region\cite{3Ohashi3}.
In the present letter, we emphasize this similarity by considering the
single-particle Fermi quasiparticle spectrum in the BCS-BEC crossover region 
with a Feshbach resonance. We point out that the well-known BCS-Bogoliubov
excitation spectrum is a good approximation in both the BCS and BEC
limits\cite{3Ohashi3} at all temperatures in the superfluid phase. In
particular, the Fermi excitations exhibit an
energy gap due to their coupling to a molecular Bose condensate, the
analogue of what happens in the classic BCS limit. Our work emphasizes
the importance of measuring\cite{12Torma} the energy and momentum of
single-particle
excitations in trapped superfluid Fermi gases as a way of
studying the effect of the molecular Bose condensate.
We do not address possible corrections very 
close to the Feshbach resonance in the unitarity limit 
($k_{\rm F}a^{\rm 2b}_s\to\pm\infty$).
\par
We consider a gas of Fermi atoms composed of two atomic hyperfine
states (labeled by $\sigma=\uparrow,\downarrow$), coupled to a
molecular two-particle bound state. The coupled fermion-boson model
Hamiltonian is given
by\cite{1Ohashi1,7Holland,8Timmermans,13Ranninger} 
\begin{eqnarray}
{\hat H}
&=&
\sum_{{\sib p}\sigma}\varepsilon_{\sib p}
c^\dagger_{{\sib p}\sigma}c_{{\sib p}\sigma}
+
\sum_{{\sib q}}(\varepsilon_{B{\sib q}}+2\nu)
b^\dagger_{{\sib q}}b_{{\sib q}}
\nonumber
\\
&-&U\sum_{{\sib p},{\sib p}',{\bf q}}
c^\dagger_{{\sib p}+{\bf q}/2\uparrow}c^\dagger_{-{\sib p}+{\bf q}/2\downarrow}
c_{-{\sib p}'+{\bf q}/2\downarrow}c_{{\sib p}'+{\bf q}/2\uparrow}
\nonumber
\\
&+&
g_{\rm r}\sum_{{\sib p},{\sib q}}
[
b^\dagger_{{\sib q}}
c_{-{\sib p}+{\sib q}/2\downarrow}c_{{\sib p}+{\sib q}/2\uparrow}
+{\rm h.c.}
].
\label{eq.1}
\end{eqnarray}
For simplicity, we discuss the case of a {\it uniform} two-component Fermi
gas, although results in the figures take into account an isotropic
harmonic trap potential. In Eq. (\ref{eq.1}), a
Fermi atom and a Boson associated with the Feshbach
resonance are, respectively, described by the destruction operators
$c_{{\bf p}\sigma}$ and $b_{\bf q}$. The kinetic energy of a bare
Fermi atom is $\varepsilon_{\bf p}\equiv p^2/2m$ and
$\varepsilon_{B{\bf q}}+2\nu\equiv q^2/2M+2\nu$ is the excitation spectrum
of the bare $b$-molecular bosons. The lowest energy of the
$b$-bosons ($2\nu$) is the threshold
energy of the Feshbach resonance. The last term in Eq. (\ref{eq.1})
is the Feshbach resonance with a coupling constant $g_{\rm r}$,
which describes how a $b$-molecule can dissociate into two Fermi atoms
and how two Fermi atoms can form one $b$-boson. Eq. (\ref{eq.1})
also includes the usual attractive interaction $-U~(<0)$
arising from nonresonant processes.
\par
Since a $b$-Bose molecule consists of a bound state of two Fermi atoms, the boson mass is $M=2m$ and the conservation of the total number of particles $N$ imposes the relation
\begin{equation}
N=\sum_{{\bf p}\sigma}
\langle c_{{\bf p}\sigma}^\dagger c_{{\bf p}\sigma}\rangle
+
2\sum_{\bf q}
\langle b_{\bf q}^\dagger b_{\bf q}\rangle.
\label{eq.2}
\end{equation}
We incorporate this crucial constraint into the model Hamiltonian in
Eq. (\ref{eq.1}) using a single chemical potential and work with 
${\cal H}\equiv H-\mu
N$. This leads to a shift in the energies,
$\varepsilon_{\bf p}\to\xi_{\bf
  p}\equiv\varepsilon_{\bf p}-\mu$ and $\varepsilon_{B{\bf
    q}}+2\nu\to\xi_{B{\bf q}}\equiv\varepsilon_{B{\bf q}}+2\nu-2\mu$.
\par
The mean-field approximation (MFA) for the coupled
Fermi-Bose Hamiltonian reduces to\cite{13Ranninger,note10}
\begin{eqnarray}
{\cal H}_{\rm MFA}
&=&
\sum_{{\bf p}\sigma}\xi_{\bf p}c^\dagger_{{\bf p}\sigma}c_{{\bf
    p}\sigma}
-\sum_{\bf p}[\Delta c_{{\bf p}\downarrow}c_{-{\bf p}\uparrow}+h.c.]
\nonumber
\\
&+&\sum_{{\bf q}\ne0}\xi_{B{\bf q}}b^\dagger_{\bf q}b_{\bf q}
+
g_{\rm r}\sum_{\bf p}[\phi_m c_{{\bf p}\downarrow}c_{-{\bf p}\uparrow}+h.c.],
\label{eq.3}
\end{eqnarray}
where the Cooper pair order parameter is $\Delta\equiv U\sum_{\bf
  p}\langle c_{-{\bf p}\downarrow}c_{{\bf p}\uparrow}\rangle$ and the
molecular condensate is described by $\phi_m\equiv \langle b_{{\bf q}=0}\rangle
$. Thus the MFA broken symmetry state is described by
\begin{eqnarray}
{\cal H}_{\rm MFA}
&=&
\sum_{{\bf p}\sigma}\xi_{\bf p}c^\dagger_{{\bf p}\sigma}c_{{\bf p}\sigma}
+\sum_{{\bf q}\ne 0}\xi_{B{\bf q}}b^\dagger_{\bf q}b_{\bf q}
\nonumber
\\
&-&
\sum_{\bf p}[{\tilde \Delta} c_{{\bf p}\downarrow}c_{-{\bf p}\uparrow}+h.c.],
\label{eq.5}
\end{eqnarray}
where the {\it composite} order parameter is given by ${\tilde \Delta}
=\Delta-g_{\rm r}\phi_m$.
The molecular Bose condensate $\phi_m$ and the Cooper pair order
parameter $\Delta$ are strongly hybridized. 
One can also show
that\cite{1Ohashi1,3Ohashi3,Milstein,7Holland,8Timmermans}
\begin{equation}
\phi_m=-{g_{\rm r} \over 2\nu-2\mu}
\Bigl({\Delta \over U}\Bigr),
\label{eq.7}
\end{equation}
which leads to a renormalized pairing
interaction $U_{\rm eff}$
\begin{equation}
{\tilde \Delta}=
U_{\rm eff}
\sum_{\bf p}
\langle
c_{-{\bf p}\downarrow} c_{{\bf p}\uparrow}
\rangle,~~~
U_{\rm eff}=U+
{g_{\rm r}^2 \over 2\nu-2\mu}.
\label{eq.8}
\end{equation}
\par
In recent discussions on the BCS-BEC crossover (for example,
Refs. \cite{Viverit,Combescott}), 
the Feshbach resonance is often only included to the
extent that the order parameter ${\tilde \Delta}$ in Eq. (\ref{eq.8})
involves the correct two-body $s$-wave scattering length given by
$-4\pi a_s^{\rm 2b}\hbar^2/m=U+g_r^2/2\nu$.
The two-body Feshbach resonance at $2\nu=0$ occurs at the resonance
value $B_0$ of the magnetic field. Such calculations are 
identical to the original
theories of the BCS-BEC crossover in
superconductors\cite{4Leggett,Melo,Randeria}. No explicit account is
taken of the formation of a $b$-molecule condensate described by
$\phi_m$, as done in the above MFA calculation, which
leads to the effective pairing interaction in
Eq. (\ref{eq.8})\cite{1Ohashi1,3Ohashi3,Milstein}.
\par
Using the simple pairing MFA summarized by
Eqs. (\ref{eq.5})-(\ref{eq.8}) gives the usual BCS-Gor'kov expressions for 
the diagonal and off-diagonal single-particle Green's functions
\begin{eqnarray}
G_{11}({\bf p},\omega)
=
{\omega+\xi_{\bf p} \over \omega^2-E_{\bf p}^2},
~~~G_{12}({\bf p},\omega)
=
-{{\tilde \Delta} \over \omega^2-E_{\bf p}^2},
\label{eq.10}
\end{eqnarray}
with the BCS-Bogoliubov excitation energy given by
\begin{equation}
E_{\bf p}=\sqrt{(\varepsilon_{\bf p}-\mu)^2+|{\tilde \Delta}|^2}.
\label{eq.11}
\end{equation}
Using $G_{12}({\bf p},\omega)$ to calculate $\Delta$, 
one finds that ${\tilde \Delta}$
satisfies the ``gap equation" with the pairing interaction $U_{\rm eff}$,
\begin{equation}
{\tilde \Delta}=U_{\rm eff}
\sum_{\bf p}
{{\tilde \Delta} \over 2E_{\bf p}}\tanh{1 \over 2}\beta E_{\bf p}.
\label{eq.13}
\end{equation}
As usual, a cutoff is introduced in the momentum
summation\cite{1Ohashi1,Milstein}. At 
this MFA level, the total number of atoms at $T=0$ is 
\begin{equation}
N=N_{\rm F}+2N^{\rm c}_{\rm B},
\label{eq.13b}
\end{equation}
where the number of Fermi atoms is given [using $G_{11}({\bf p},\omega)$]
by the well-known expression
\begin{equation}
N_{\rm F}=
\sum_{{\bf p},\sigma}
\langle c_{{\bf p}\sigma}^\dagger c_{{\bf p}\sigma}
\rangle
=
\sum_{\bf p}
\Bigl[
1-{\xi_{\bf p} \over E_{\bf p}}\tanh{1 \over 2}\beta E_{\bf p},
\Bigr].
\label{eq.12}
\end{equation}
The number of Bose-condensed $b$-molecules is $N^{\rm c}_{\rm B}=|\phi_m|^2$,
where $\phi_m$ is determined by ${\tilde \Delta}$ and $\mu$.
\par
We briefly sketch how one proceeds in the case 
of an isotropic harmonic trap\cite{note2}. 
We expand the fermion quantum field
operator ${\hat \Psi}_\sigma({\bf r})$ and boson operator 
${\hat \Phi}({\bf r})$ in terms of the single-particle
eigenfunctions $f_{nlm}({\bf r})\equiv u_{nl}(r)Y_{lm}(\theta,\phi)$ 
of the harmonic potential, with energy 
$\xi_{nl}=(2n+l+3/2)\hbar\omega_0$. The resulting MFA
Hamiltonian, which corresponds to Eq. (\ref{eq.5}) for a uniform gas,
is given by\cite{note2b}
\begin{eqnarray}
{\cal H}_{\rm MFA}
&=&
\sum_{nlm,\sigma}\xi_{nl}c_{nlm\sigma}^\dagger c_{nlm\sigma}
+
\sum_{nlm}\xi^B_{nl}b_{nlm}^\dagger b_{nlm}
\nonumber
\\
&-&\sum_{nn'lm}F_{nn'}^l[c^\dagger_{nlm\uparrow}c^\dagger_{n'l-m\downarrow}+h.c.].
\label{eq.ap1}
\end{eqnarray}  
Here
$c_{nlm}^\dagger$ ($b_{lmn}$) is the creation 
operator of an atom (molecule) in the state
$f_{nlm}({\bf r})$ ($f^B_{nlm}({\bf r})$). $F_{nn'}^l$ arises from the pair potential,
$F_{nn'}^l\equiv \int_0^\infty r^2dr{\tilde
  \Delta}(r)u_{nl}(r)u_{n'l}(r)$.
Eq. (\ref{eq.ap1}) can be diagonalized by the usual BCS-Bogoliubov
transformation. One can calculate 
$\Delta(r)=U\langle{\hat \Psi}_\downarrow({\bf
  r}){\hat \Psi}_\uparrow({\bf r})\rangle$ as well as $N_{\rm F}$ in
Eq. (\ref{eq.13b}), by expanding the fermion operators in terms of
the BCS-Bogoliubov excitation operators. Using the generalization of 
Eq. (\ref{eq.7}),
\begin{equation}
{g_{\rm r} \over U}\Delta(r)+
\Bigl(
-{\hbar^2\nabla^2 \over 2M}+{1 \over 2}M\omega_0^2r^2+2\nu-2\mu
\Bigr)\phi_m(r)=0,
\label{eq.ap3}
\end{equation}
we can determine $\phi_m(r)$ from $\Delta(r)$, and hence find 
the composite order parameter ${\tilde \Delta}(r)$ in $F_{nn'}^l$, as
well as $N^{\rm c}_{\rm B}$. 
\par
The values of $\mu$ and ${\tilde \Delta}$ are determined by the
self-consistent solutions of the MFA gap equation ({\ref{eq.13}) and
  the number equation given by Eqs. (\ref{eq.13b}) and (\ref{eq.12}).
This simple ``pairing approximation" for the single-particle
excitations 
is expected to give a quantitative description at $T=0$ (where $\tanh
{1 \over 2}\beta E_{\bf p}\to 1$) since
fluctuations are small and all the $b$-molecules are
Bose-condensed. This $T=0$ limit was first studied by
Leggett\cite{4Leggett} in the absence of a Feshbach resonance. 
For $T$ approaching $T_{\rm c}$, however, the fluctuations
associated with exciting $b$-molecules out of the condensate and
coupling to the particle-particle (Cooper-pair) channel become
increasingly important\cite{1Ohashi1,Milstein}. These
fluctuations (rather than the breaking up of two-particle bound
states) determine $T_{\rm c}$ and the region above, 
as first discussed by Nozi\`eres and Schmitt-Rink
(NSR)\cite{5Nozieres}. Ref. \cite{1Ohashi1} generalized the NSR
approach to deal with a Feshbach resonance by including the contribution 
of $b$-molecules {\it outside} the
condensate (ignored in Eq. (\ref{eq.13b})) in determining 
$\mu$ at $T_{\rm c}$.
\par
In Ref. \cite{3Ohashi3}, we extended the NSR approach to discuss the
superfluid phase {\it below} $T_{\rm c}$ in a uniform Fermi gas. 
Both ${\tilde \Delta}$ and $\mu$ are obtained by solving 
Eqs. (\ref{eq.13}) and (\ref{eq.13b})
self-consistently, but now including the depletion of ${\tilde \Delta}$
through the presence of non-Bose-condensed $b$-molecules\cite{15note}.
This procedure gives the simplest ``renormalization'' of the MFA-BCS
single-particle results in Eqs. (\ref{eq.10}) and (\ref{eq.11}), which
now involve values of $\mu$ and ${\tilde \Delta}$ which include the
effect of fluctuations around the MFA theory. The physics of this 
is most clearly shown in a functional
integral treatment\cite{Ohashiz}. The gap equation (\ref{eq.13}) determines the MFA
saddle point minimum, while the new number equation (\ref{eq.13b})
describes the effect of Gaussian fluctuations around the MFA saddle
point. The latter gives rise to a renormalized value of the chemical
potential $\mu$ occurring in the MFA gap equation, which in turn 
leads to a renormalized 
value of ${\tilde \Delta}$.
It is important to note that this renormalized pairing
approximation generates response
functions\cite{3Ohashi3}
 which exhibit 
gapless Goldstone modes, a crucial requirement of a conserving
many-body approximation. 
Here we 
use the $T=0$ results to illustrate the behavior
in a trapped Fermi gas\cite{note2}.

%%%%%%%%%%%%%%%%%%%%%%%%%%%%%%%%%%%%%%%%
\begin{figure}
\includegraphics[width=7cm,height=4.7cm]{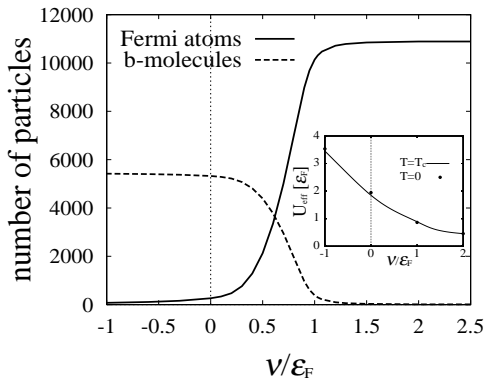}% 
\caption{\label{fig1} 
The number of particles as a function of the 
threshold energy $2\nu$ in a trapped superfluid Fermi gas at
$T=0$. The Fermi energy $\varepsilon_{\rm F}=31.5\hbar\omega_0$
is for a free gas with $N=10912$ atoms. Figs. 1-3 are for
$U=0.52\varepsilon_{\rm F}$ and $g_{\rm r}=0.2\varepsilon_{\rm F}$.
The inset shows
$U_{\rm eff}$ in a {\it uniform} Fermi gas, for 
$U=0.3\varepsilon_{\rm F}$ and $g_{\rm r}=0.6\varepsilon_{\rm F}$. 
$\mu$ is almost temperature
independent below $T_{\rm c}$, as is $U_{\rm eff}$.
}
\end{figure}
%%%%%%%%%%%%%%%%%%%%%%%%%%%%%%%%%%%%%%%%%

\par
In Fig. 1, we illustrate
how $N_{\rm F}$ and $N_{\rm B}$ vary as a function of the $b$-molecule
threshold $2\nu$ for a trapped
Fermi gas at $T=0$\cite{note2}. The molecular condensate region
\cite{9Jin,10Ketterle,11Jochim} occurs just below the two-body
Feshbach resonance at $2\nu=0$, where $a^{\rm 2b}_s$ is large and positive, leading
to stable (long-lived) $b$-molecules. In Fig. 1, the Fermi
contribution includes the Cooper pairs. The latter contribution
becomes less important for $2\nu<2\varepsilon_{\rm F}$ and is almost
absent relative to the $b$-molecules as soon as one passes through the
Feshbach resonance (when $\nu<0$, one has $a^{\rm 2b}_s>0$). A more detailed
decomposition into Fermi excitations,
Cooper-pairs and $b$-molecules is given at $T_{\rm c}$ in
Ref. \cite{1Ohashi1}, which may be more relevant to the region just
below $T_{\rm c}$ studied in Refs. \cite{9Jin,10Ketterle,11Jochim}.
\par

%%%%%%%%%%%%%%%%%%%%%%%%%%%%%%%%%%%%%%%%%%%%
\begin{figure}
\includegraphics[width=7cm,height=7.3cm]{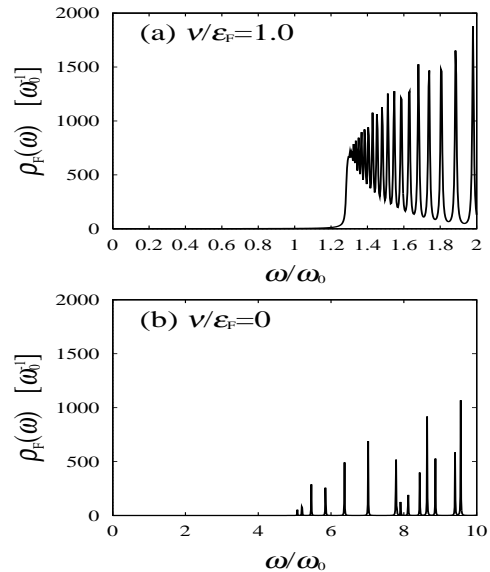}% 
\caption{\label{fig2} 
Single-particle spectral density in a trapped gas of Fermi
atoms at $T=0$. The sharp peaks reflect the discrete BCS-Bogoliubov
energies in a harmonic potential.
}
\end{figure}
%%%%%%%%%%%%%%%%%%%%%%%%%%%%%%%%%%%%%%%%%%%%

In our pairing approximation, the effective attractive
interaction $U_{\rm eff}$ in Eq. (\ref{eq.8}) plays a crucial
role. 
One finds that $U_{\rm eff}$ smoothly increases 
in magnitude as $\nu$ decreases, with 
nothing unusual happening at the two-body Feshbach resonance at
$2\nu=0$ (see the inset in Fig. 1).
\par
As discussed in Ref. \cite{3Ohashi3}, the BCS-Bogoliubov quasiparticle
spectral density is easily calculated from
\begin{equation}
\rho_{\rm F}(\omega)\equiv -{1 \over \pi}\sum_{\bf p}
{\rm Im}
[G_{11}({\bf p},\omega+i0^+)].
\label{eq.14}
\end{equation}
The spectral density for a uniform gas at $T=0$ is shown in Fig. 8 
in Ref. \cite{3Ohashi3}. In the BEC region $2\nu\lesssim 0$,
we have $\mu<0$ and as a result, 
$\rho_{\rm F}(\omega)$ has an 
excitation energy gap which starts 
at $\omega=\sqrt{\mu^2+|{\tilde \Delta}|^2}$ 
for excitations 
with ${\bf p}=0$\cite{4Leggett,Randeria}. 
In the BCS region $2\nu>0$ (where $\mu>0$), the quasiparticle gap
starts at $\omega=|{\tilde \Delta}|$, from Fermi quasiparticles with momentum
$p=\sqrt{2m\mu}$. In Fig. 2, we
compare the density of states $\rho_{\rm F}(\omega)$ at
$\nu=\varepsilon_{\rm F}$ (BCS) with the result at the Feshbach
resonance $\nu=0$ (BEC) in a trap. The sharp peaks correspond to the
BCS-Bogoliubov excitation frequencies obtained from numerically
solving\cite{note2} the Bogoliubov-de Gennes coupled equations,
together with a  self-consistent calculation of $\mu$ and ${\tilde
  \Delta}({\bf r})$ as described above. 
The sharp excitation edge in Fig. 2(a) reflects the
BCS density of states (DOS) $\rho_{\rm F}(\omega)\sim
Re[{\omega \over \sqrt{\omega^2-|{\tilde \Delta}|^2}}]$, which is 
absent in the BEC region shown in Fig. 2(b). 
\par
The threshold energy of the continuum spectrum in the density
response function\cite{3Ohashi3} is equal to the minimum 
energy needed to break a
bound pair of atoms and put them into the lowest energy states.
This is the binding energy $E_{\rm BE}$ of the dimer in the
interacting Fermi gas. 
In the JILA experiments\cite{9Jin,Regal}, a r.f. pulse of energy
$h\nu_{\rm r.f}$ is used to stimulate a transition to a lower Zeeman
state. The two atoms will share an excess energy $\Delta E=
h\nu_{\rm atom}-h\nu_{\rm r.f}-E_{\rm BE}$, where $h\nu_{\rm atom}$ is
the atom-atom transition frequency. 
If the dissociated pair of atoms
are in the $\uparrow$ ($m_{\rm F}=-9/2$) and $\downarrow$ 
($m_{\rm F}=-7/2$) states involved in the molecular Bose condensate,
the energy of each atom ($E_{\rm atom}$) will have a threshold $E_{\rm 
  g}$ given by $E_{\rm atom}\ge E_{\rm g}
=\sqrt{\mu^2+|{\tilde \Delta}|^2}-|\mu|$ in the BEC region ($\mu<0$),
with the atoms having very small momentum. For ${\tilde
  \Delta}\ll|\mu|$, this gap $E_{\rm g}\simeq |{\tilde
  \Delta}|^2/2|\mu|$ 
is very small. The excess energy $\Delta E$ will be shared by the
dissociated atoms, with $E_{\rm atom}=E_{\rm g}+{\Delta E \over 2}$.
In contrast, one has $E_{\rm g}=|{\tilde \Delta}|+\mu$ in the BCS
region ($\mu>0$), with the atoms having a large momentum ($p=\sqrt{2m\mu}$)
comparable to the Fermi momentum ($\mu\sim\varepsilon_{\rm
  F}\gg|{\tilde \Delta}|$). In Fig. 3, we plot $E_{\rm g}$ as a
function
of $\nu$ in a trapped Fermi gas at $T=0$.   
Of course, these predictions assume that the dissociated atoms have 
thermalized with the rest of the atoms in the Fermi gas, so that the
excitation spectrum is described by the equivalent of Eq. (\ref{eq.11}).
\par
%%%%%%%%%%%%%%%%%%%%%%%%%%%%%%%%%%%%%%%%%%%%%%%
\begin{figure}
\includegraphics[width=7cm,height=4.7cm]{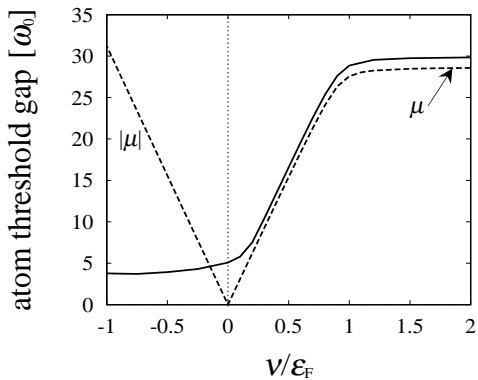}% 
\caption{\label{fig3} 
Energy gap for atoms vs $\nu$ in a superfluid Fermi gas at
$T=0$ trapped in a harmonic potential. 
}
\end{figure}
%%%%%%%%%%%%%%%%%%%%%%%%%%%%%%%%%%%%%%%%%%%%%%%
By way of contrast, if the r.f. pulse dissociates the molecule into
atoms in the spin state $\uparrow$ ($m_{\rm F}=-9/2$) and another
Zeeman state ($m_{\rm F}=-5/2$ in Ref. \cite{9Jin}), only the
$\uparrow$ atom will be coupled to the underlying molecular
condensate involving ($\uparrow,\downarrow$)-pairs. This
atom will exhibit an energy gap $E_{\rm g}$ associated with the
finite value of the order parameter $|{\tilde \Delta}|$.
The other atom will not have an energy gap. One sees that the energy of
the pair of dissociated atoms can be different, assuming that a
molecular condensate is only coupled to one of them. 
The energy and momentum spectrum of dissociated atoms
should be a very direct probe of the presence (or absence) of a Bose
condensate (see also Refs.\cite{12Torma,Viverit} for other techniques to
study single-particle excitations).
\par
In summary, we recall that in the original work on the BCS-BEC
crossover\cite{4Leggett,5Nozieres,Melo,Randeria}, the only bound states were
Cooper pairs, whose existence is a many-body effect and does not
require a two-body resonance. In this case, when $a_s>0$, the
Cooper pairs are long-lived and form a molecular condensate. In
contrast, in the presence of a Feshbach resonance, the condensate is
formed of stable dimer molecules in the BEC limit, 
with a negligible contribution from
Cooper pairs. We
have studied the single-particle quasiparticle spectrum associated
with a {\it composite}
Bose order parameter arising from the formation of a
BEC of Cooper pairs and Feshbach-induced molecules. We have emphasized
that a simple static MFA pairing approximation renormalized to include
fluctuations\cite{3Ohashi3,5Nozieres}
describes {\it both} the BCS limit (where Cooper
pairs dominate) and the BEC limit (where stable Feshbach molecules
dominate). Both regions are
described by a renormalized BCS-type single-particle Fermi spectrum. 
\par
Y. O. was supported by a University of Tsukuba Research Project and
A. G. by NSERC of Canada. 

%
%%%%%%%%%%%%%%%%%%%%%%%%%%%%%%%%%%%%%%%%%%%%%%%%%%%%%%%%%%%%%%%%%%%%%

%%%%%%%%%%%%%%%%%%%%%%%%%%%%%%%%%%%%%%%%%%%%%%%%%%%%%%%%%%%%%%%%%%%%%%%%%%%%%%
%


\begin{references}
\bibitem{note10} We omit the Bose condensate term $(2\nu-2\mu)|\phi_m|^2$.
\bibitem{note2b} We also include the Hartree term in our calculations.
\end{references}
\end{document}